\documentclass[twocolumn]{revtex4}

\usepackage{amsmath}
\usepackage{graphicx}
\begin{document}

\title{Measuring the speed of light using beating longitudinal modes in an open-cavity HeNe laser}

\author{Daniel J. D'Orazio}
\altaffiliation{A 2009 graduate of Juniata College currently supported by a Fulbright Fellowship at the University of Z\"urich.}
\author{Mark Pearson}
\author{Justin T. Schultz}
\altaffiliation{A 2008 graduate of Juniata College and 2008--2009 Fulbright Fellow at the Australian National University. He is now doing graduate work at the Optics Institute, University of Rochester.}

\author{Daniel Sidor}
\altaffiliation{A 2007 graduate of Juniata College currently doing graduate work at the Optics Institute, University of Rochester.}

\author{Michael Best}
\altaffiliation{A 2006 graduate of Juniata College currently doing graduate work at the University of Pittsburgh School of Medicine.}

\author{Kenneth Goodfellow}
\author{Robert E. Scholten}
\altaffiliation{ARC Centre of Excellence for Coherent X-ray Science, School of Physics, The University of Melbourne, Victoria 3010, Australia.}

\author{James D. White}
\email{white@juniata.edu}
\affiliation{Juniata College, Department of Physics, Huntingdon, Pennsylvania 16652}

\begin{abstract}
We describe an undergraduate laboratory that combines an accurate measurement of the speed of light, a fundamental investigation of a basic laser system, and a nontrivial use of statistical analysis. Students grapple with the existence of longitudinal modes in a laser cavity as they change the cavity length of an adjustable-cavity HeNe laser and tune the cavity to produce lasing in the TEM$_{00}$ mode. For appropriate laser cavity lengths, the laser gain curve of a HeNe laser allows simultaneous operation of multiple longitudinal modes. The difference frequency between the modes is measured using a self-heterodyne detection with a diode photodetector and a radio frequency spectrum analyzer. Asymmetric effects due to frequency pushing and frequency pulling, as well as transverse modes, are minimized by simultaneously monitoring and adjusting the mode structure as viewed with a Fabry-Perot interferometer. The frequency spacing of longitudinal modes is proportional to the inverse of the cavity length with a proportionality constant equal to half the speed of light. By changing the length of the cavity, without changing the path length within the HeNe gas, the speed of light in air can be measured to be ($2.9972 \pm0.0002) \times 10^{8}$ m/s,  which is to high enough precision to distinguish between the speed of light in air and that in a vacuum.
\end{abstract}

\maketitle

\section{Introduction}
\begin{figure}[h!]
\centering
\includegraphics[width=0.37\textwidth]{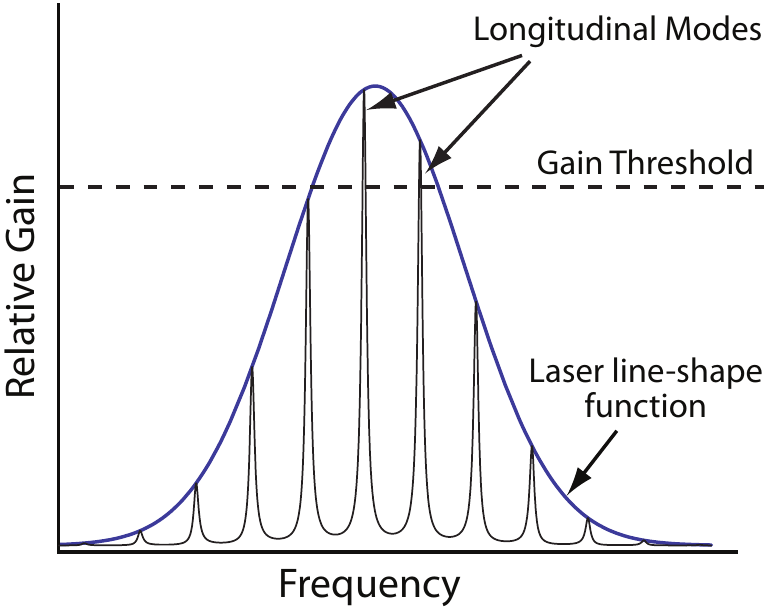}
\caption{Schematic illustration of the longitudinal cavity modes and gain bandwidth of a laser. In the situation shown, the net gain minus losses is sufficient for laser output at only two longitudinal cavity modes. The beat frequency that we observe to measure the speed of light is the spacing between these adjacent modes.}
\label{fig:long_modes2}
\end{figure}

Lasers emit light over a range of wavelengths
described by the laser line shape
function.\cite{Csele,Milonni,Pedrotti} For a HeNe laser operating
under normal conditions, the main source of laser line shape
broadening is Doppler broadening in the lasing medium, resulting in
a Gaussian gain profile (see Fig.~\ref{fig:long_modes2}). The laser does not
emit a continuous spectrum of wavelengths over this Gaussian
gain-permitted wavelength range; rather, it can only lase when there
is resonance in the lasing cavity. For the TEM$_{00}$ mode there
exists an integer number, $N$, of half wavelengths between the mirrors
of the laser cavity, resulting in the allowed resonance wavelengths
\begin{equation}
\lambda_N = \frac{2nL}{N},
\end{equation}
where $L$ is the length of the laser cavity and $n$ is the index of refraction of the medium filling the laser cavity.

The laser output consists of discrete wavelength peaks with power dictated by the Gaussian line shape envelope and the unsaturated gain threshold (see Fig.~\ref{fig:long_modes2}). These peaks are called longitudinal cavity modes. When the laser cavity supports more than one peak (that is, where the gain is greater than the losses for those peaks), the laser output consists of multiple discrete wavelengths. If the light from these multiple modes is projected onto a detector (for example, a photodiode), then the photocurrent will oscillate at the difference frequency, producing a beat signal.

The beat frequency of interest is at the frequency due to the spacing between adjacent longitudinal modes. The frequency of the $N$th mode can be derived from Eq.~(1) to be $f_N = N(c/2nL)$. Thus the beat frequency is given by
\begin{equation}
\Delta f = {c\over{\lambda_{N+1}}} - {c\over{\lambda_{N}}} = {c \over 2nL},
\end{equation}
and therefore $L = c/2n\Delta f$, indicating that the cavity length is directly proportional to the reciprocal of the beat frequency.\cite{Razdan} Observing the variation in beat frequency between adjacent longitudinal modes with the cavity length $L$ gives the speed of light.

Accurate measurements of the beat frequency are accomplished
inexpensively by directing the output of the laser onto a
high-speed photodetector\cite{Detectors} monitored with an RF
spectrum analyzer or frequency counter.\cite{CSA,Phillips,Conroy} This
approach has been demonstrated in Ref.~\onlinecite{Brickner} in an undergraduate experiment with the goal of measuring the speed of light using the relation in Eq.~(2) for a
single laser cavity length and single corresponding beat
frequency. The method is easily understood because it is analogous to investigations of waves on a string. It has a drawback, however; the inability to obtain a precise
measurement of the cavity length (from the inner-cavity
side of the output coupler to inner-cavity side of the back mirror)
inevitably leads to results that are only marginally better than those
obtained with standard time-of-flight or Foucault methods commonly
used in undergraduate physics laboratories, which typically yield
measurements accurate to within $\approx \pm 1 \%$.\cite{Bates,Fiber,Foucault} Minor
improvements on this method can be made by collecting
data for multiple lasers of different lengths and plotting the beat
frequency as a function of cavity length. In addition to the
uncertainty in length between the mirrors, there is also the problem
of not knowing a precise (and constant) value for the index of
refraction inside the gas tube. These
obstacles can be overcome by using the laser as a simple light source,
amplitude modulated at the intermode beat frequency, and measuring the phase
difference between detectors placed at two different locations along
the laser path.\cite{Barr} This modulation technique improves the measurement of
the speed of light by an order of magnitude, but
at the cost of increasing the conceptual complexity. The introduction of the
adjustable-length HeNe laser significantly reduces the consequences of
uncertainty in mirror location and the index of refraction, and improves the
measurement by a order of magnitude over the modulation technique,
while retaining the conceptual simplicity of the original
study of Ref.~\onlinecite{Brickner}.

\section{Methods}
\begin{figure}[h!]
\centering
\includegraphics[width=0.37\textwidth]{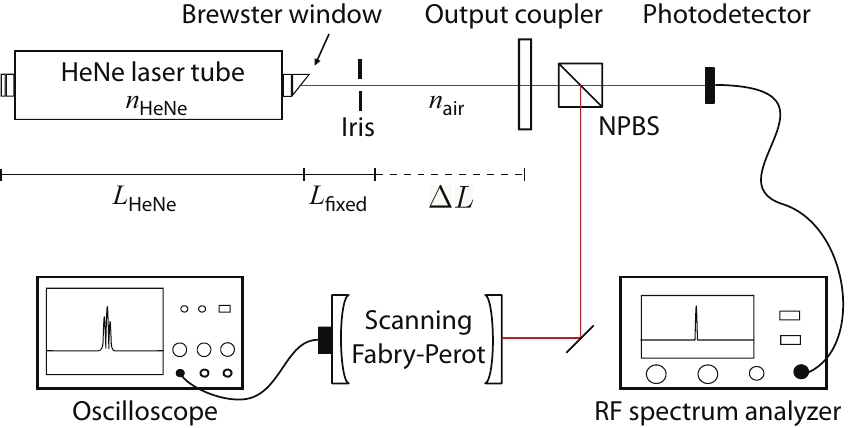}
\caption{A schematic of the experimental set-up. The length of the cavity can be adjusted over a range of approximately 16\,cm by sliding the output coupler along an optical track. The mode structure of the laser output is monitored using a scanning Fabry-Perot interferometer with a free spectral range of 1.5\,GHz and a Finesse of 250. The mode structure is controlled via an adjustable iris in the cavity. The portion of the beam that is not analyzed by the Fabry-Perot is incident on a fast photodetector (1\,ns rise time), which is coupled to an RF spectrum analyzer on which the beat signal between adjacent longitudinal modes is observed. (NPBS = non-polarizing beam splitter.)}
\label{fig:HeNe_set-up}
\end{figure}

Figure~\ref{fig:HeNe_set-up} represents a schematic of the
experimental set-up. The laser has an adjustable, open-cavity design
with a 28\,cm HeNe plasma tube terminated on one side with a mirror and
on the other with a Brewster window. The Brewster window suppresses
modes with polarization orthogonal to the Brewster plane, so that all
supported modes have the same polarization and thus mix
effectively in the photodetector.\cite{Csele, Milonni}
The experiment can be conducted without a Brewster window, but due to mode competition, adjacent longitudinal modes
are typically polarized orthogonal to each other and do not mix
in the photodetector, resulting in an observed signal with twice the
expected frequency.\cite{Tang} If a Brewster window is not present,
the situation can be remedied by placing a linear polarizer in front
of the photodetector to project the polarizations of adjacent modes
onto a common axis.

The variable-length cavity system has been reported and
widely used in undergraduate labs to
explore laser cavity modes and
stability.\cite{Brandenberger,Polik,Jackson,Melles}
The output coupler is a 0.60\,m radius-of-curvature mirror held in
a gimbal mount. It is attached to a
sliding track, allowing the cavity length to be changed from
$\approx 38$\,cm (lower bound limited by the length of the plasma tube) up to $\approx 54$\,cm (upper bound restricted by laser losses). Typically we see two or three longitudinal modes separated by about 300\,MHz within the 1.5\,GHz gain bandwidth of the HeNe medium.\cite{Pedrotti} Inside the
cavity, between the output coupler and the plasma tube, is an iris used to restrict gain in the region away from the
optical axis of the cavity and thus force the laser to emit in the
TEM$_{00}$ (Gaussian) mode. Restricting the laser to a single transverse mode is necessary because higher-order modes produce additional beat frequencies that complicate the RF
spectrum. The allowed frequencies for the TEM$_{00}$ mode are given
by Eq.~(1), and the allowed frequencies for higher-order
TEM$_{ij}$ modes are given by
\begin{equation}
f_{Nij} = {c \over{2L}} \left[ N + {1 \over{\pi}}(i + j + 1)\cos^{-1}(\sqrt{g_1g_2}) \right],
\end{equation}
where $N$ is the same mode number as in Eq.~(1) and $g_1g_2$ is the
resonator stability.\cite{Milonni, Goldsborough} Thus if TEM$_{00}$
and TEM$_{ij}$ are allowed to exist simultaneously in the cavity, beat
frequencies will exist at ${c/2L}$ and ${c/2L} \pm {(1/\pi)} (i + j +
1)\cos^{-1}\sqrt{g_1g_2}$. These additional beat
frequencies could provide an interesting method for measuring the
resonator stability, $g_1g_2$, for a fixed cavity length.

\subsection{Cavity Length Measurement}
As noted in Sec.~I, we cannot accurately measure the entire
laser cavity length due to the uncertainty of the position of the
mirror in the HeNe tube. In addition, the index of refraction within
the He- and Ne-filled tube is different from that in the rest of the
cavity, which is filled with air (and a small length of glass at
the window). Because we do not know the index of refraction inside the
laser plasma tube, we modify Eq.~(2) by splitting $L$ into the
two main regions within the laser cavity that have different indices of
refraction. Let $n_{\mathrm{HeNe}}$ be the index of refraction inside
the laser plasma tube and $n_{\rm air}$ be the index of refraction of
air between the Brewster window and the output coupler. Then, $nL =
n_\mathrm{HeNe}L_{\mathrm{HeNe}} +
n_{\mathrm{\rm air}}L_{\mathrm{\rm air}}$, where additional fixed components such as the glass window and dielectric mirror coatings are assumed in the first term. In practice neither of these $L$
values is simple to measure accurately, and thus we split $L_{\rm air}$ further
into two arbitrary pieces (a fixed length and a measured variable
length) such that $nL = n_{\mathrm{HeNe}}L_{\mathrm{HeNe}} +
n_{\mathrm{\rm air}}[L_{\mathrm{fixed}} + \Delta L]$ (see Fig.~\ref{fig:HeNe_set-up}). We substitute this expression into
Eq.~(2) and obtain
\begin{equation}
\Delta L = {c \over{2n_{\mathrm{\rm air}} \Delta f}} - {\frac{n_{\mathrm{HeNe}}}{n_{\mathrm{\rm air}}}}L_{\mathrm{HeNe}} - L_{\mathrm{fixed}},
\end{equation}
which is the equation of a line with slope $c/
2n_{\rm air}$. Equation (4) allows us to measure the
cavity length to an arbitrarily chosen reference point fixed between
the laser plasma tube output and the output coupler. In practice we
measure $\Delta L$ from a fixed block near the sliding track to the
base of the output coupler using digital vernier calipers. The speed
of light is then found from the slope of a $\Delta L$ versus $1/\Delta f$ plot. The unknown details of $n_{\mathrm HeNe}$,
$L_{\mathrm HeNe}$, and similar terms for the glass window are
gathered in the $y$-intercept. This algebraic trick
works only when the laser is in the TEM$_{00}$ mode, and does not work if
the laser were in transverse TEM$_{Nij}$ modes (where $i$ and $j$ are
nonzero), as represented in Eq.~(3). More
elegantly, we are taking the derivative of Eq.~(2) in the region of
air where we are free to move the output coupler as shown:
\begin{equation}
\frac{dL}{d(\frac{1}{\Delta f})} = \frac{c}{2n_{\rm air}}.
\end{equation}

\subsection{Frequency Measurement}
For the range of laser cavity lengths in the set-up ($\approx
0.54$\,m to 0.38\,m), the beat frequency varies from $\approx
280$\,MHz to 390\,MHz, a change of 110\,MHz over 16\,cm. The signal from the photodetector was analyzed with an RF spectrum analyzer with a maximum span of 3\,GHz and a minimum resolution bandwidth of $10$\,Hz.\cite{Detectors,CSA} A frequency counter could in
principle be used, but would not provide insight into additional beat
frequencies from transverse mode contributions. In addition to
analyzing the laser output with the photodetector and spectrum
analyzer, we split off a portion of the laser output to a scanning
Fabry-Perot interferometer to observe its longitudinal mode
structure.\cite{Fabry} The Fabry-Perot spectrum shows the
number of modes and their amplitudes (and therefore the amplitude of
the gain curve).

The amplitude of the modes provides information on frequency pulling
and pushing, which cause small but statistically significant shifts in
the beat frequency. \textit{Frequency pulling} refers to a change in the
spacing of longitudinal modes under a gain curve resulting from the
different indices of refraction experienced by each mode. Across the
range of frequencies that lie within the laser gain curve, the index
of refraction varies steeply near the resonance transition, being
lower or higher for frequencies below or above the resonance
transition. From Eq.~(2) we see that the allowed
frequencies below the gain peak occur at higher frequencies than would
be expected and vice versa. The result is a ``pulling" of the
longitudinal modes toward the center of the gain curve, effectively
decreasing the difference frequency between the two. The amount by
which the modes are pulled together and the beat frequency is lowered
is a function of the relative intensity of the two heterodyning
modes. For a given gain curve amplitude we find that the beat
frequency varies over $\approx 30$ to 40\,kHz for the full range of
mode relative intensities, in agreement with other
studies.\cite{Lindberg}

\begin{figure}[h!]
\centering
\includegraphics[width=0.4\textwidth]{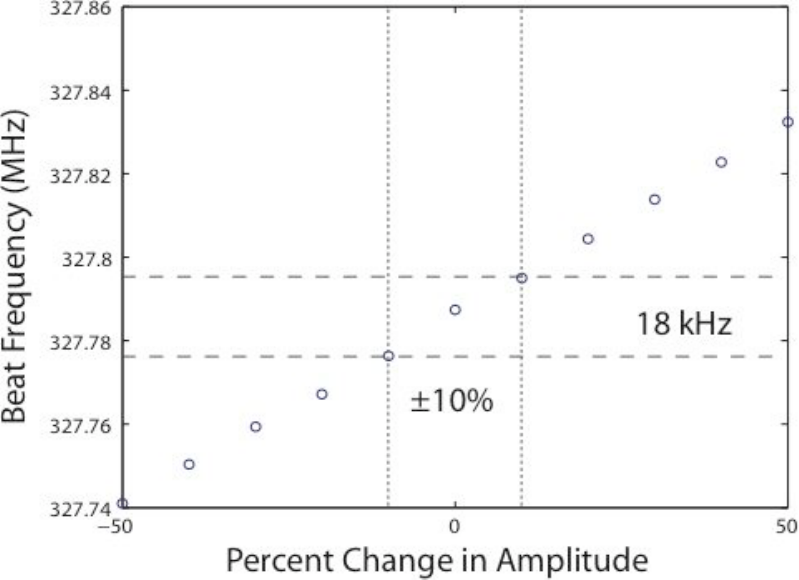}
\caption{A sample plot of beat frequency as a function of gain curve amplitude as read from the Fabry-Perot transmission showing the effects of frequency pushing. The uncertainty in the gain curve amplitude of $\pm 10\%$ corresponds to a 18\,kHz frequency variation equivalent to a $\pm 9$\,kHz uncertainty in the beat frequency. The $0\%$ mark in this figure refers to the desired amplitude at which the frequency measurement is to be taken.}
\label{fig:pushing}
\end{figure}

\textit{Frequency pushing} refers to the increase of the difference frequency
between longitudinal modes as the field intensity in the laser cavity
increases.\cite{Siegman, Shimoda} As the gain in the cavity is
increased, the beat frequency also increases. We observe this increase in our
set-up; when two adjacent longitudinal modes are observed with
identical intensities, for a $\approx 10$\% change in total amplitude of
the gain curve, there is a $\approx 9$\,kHz change in beat frequency.
Figure~\ref{fig:pushing} shows this effect over a wide range of
amplitudes, showing a linear relation between
the change in the intensity of the modes and the frequency pushing
effect. When taking data to measure the speed of light, we are
able to hold our amplitude fluctuation to a variation of $\pm 10$\%.

\begin{figure}[h!]
\centering
\includegraphics[width=0.4\textwidth]{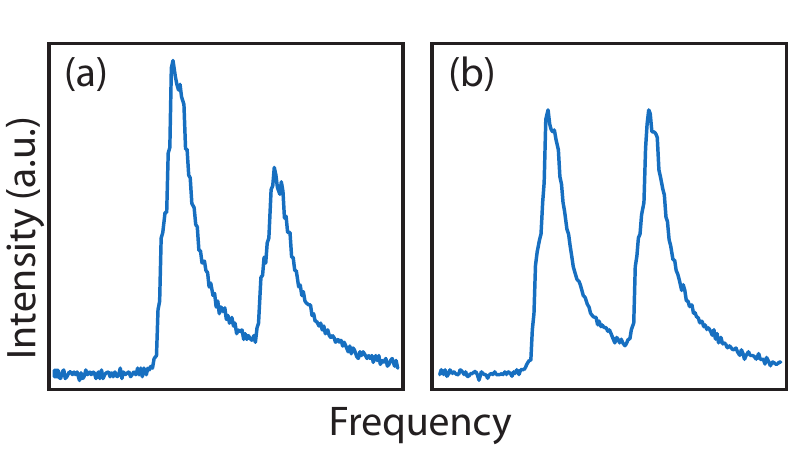}
\caption{Screen shots from the oscilloscope showing transmission of the scanning Fabry-Perot interferometer. The laser output power is the same in both cases. (a) An instance where the two mode intensities are asymmetrical around the center of the gain curve, whereas (b) shows the two modes when they have equal intensities. Due to frequency pulling, the two instances will produce beat frequency values differing by a few kHz. }
\label{fig:lmodes_neq}
\end{figure}

To minimize inconsistencies due to frequency pulling effects, we use the Fabry-Perot to ensure that each measurement (that is, the beat frequency at each cavity length) is taken for two longitudinal modes at the same relative intensities (see Fig.~\ref{fig:lmodes_neq}). The refractive index within the laser tube is then the same for both modes and very similar for all beat frequency measurements, reducing the pulling effect. More complex methods of ensuring that the two longitudinal modes are symmetric about the frequency of the emission line have been implemented in other studies.\cite{Balhorn} These involve using a non-Brewster window laser and subtracting the outputs of the orthogonal modes detected with two photodetectors and a polarizing beam splitter. This difference is used to control the electronic feedback to make slight adjustments to the length of the cavity. We have not attempted such elaborate feedback schemes. Instead, students make the necessary adjustments by applying gentle pressure to the optical table, which affects the cavity length on the micron scale.

\begin{figure}[h]
\centering
\includegraphics[width=0.4\textwidth]{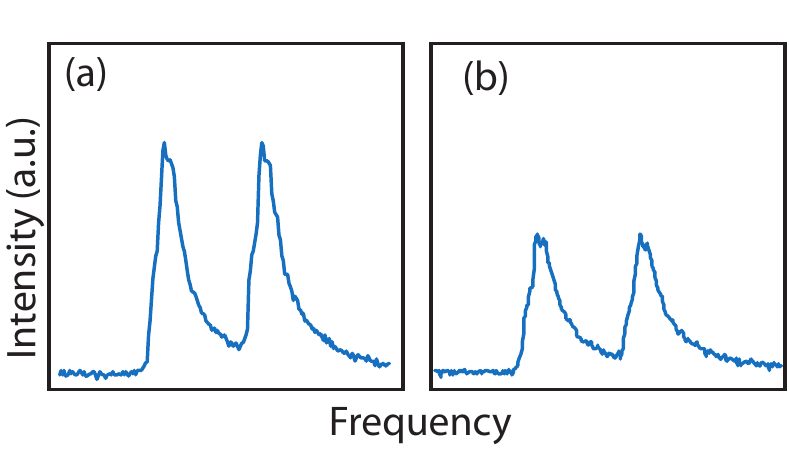}
\caption{Screen shots from the oscilloscope showing transmission of the scanning Fabry-Perot interferometer. Both show the existence of two longitudinal modes at the same relative intensity and thus each exhibit the same frequency pulling induced effects. (a) Two modes when the laser is operating at a higher gain setting than is present in (b). Due to frequency pushing, the beat frequency produced by the modes in (a) is higher than the beat frequency produced by the modes in (b). }
\label{fig:lmodes_eq}
\end{figure}

To counteract inconsistencies due to frequency pushing effects, we use the Fabry-Perot to ensure that each measurement is taken with the longitudinal modes at the same total amplitude and thus at the same laser intensity (see Fig. ~\ref{fig:lmodes_eq}). The laser power is controlled by changing the cavity loss by adjusting the intra-cavity iris.

\section{Data Analysis and Results}
\begin{figure}[h]
\centering
\includegraphics[width=0.45\textwidth]{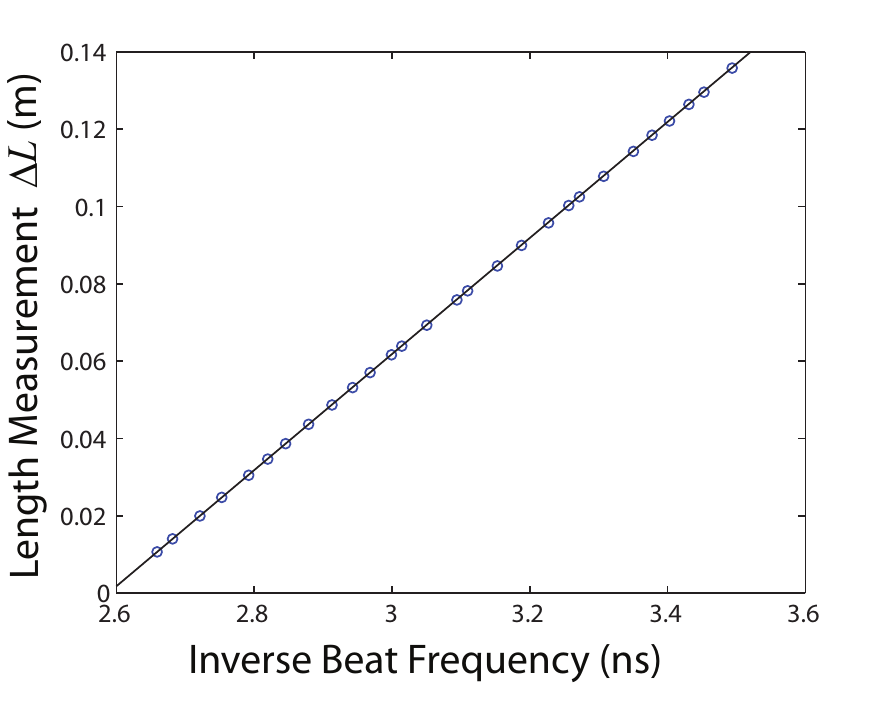}
\caption{The plot of 28 data points are fit using a weighted least squares regression. Errors are too small to display on this scale. We find a slope of $c /2n_{\rm air} = (1.4986 \pm 0.0001) \times 10^8$\,m/s.}
\label{fig:HeNe_Data_Plot}
\end{figure}

Figure~\ref{fig:HeNe_Data_Plot} represents experimental data for 28 cavity lengths. The
uncertainty in our $\Delta L$ measurement is $\pm 1 \times 10^{-5}$\,m, dictated by the measurement limit of the digital vernier calipers.
The uncertainty in our beat frequencies is dominated by frequency
variability due to frequency pulling and pushing and has been
minimized with the use of the Fabry-Perot interferometer. Due to
frequency pulling and pushing, a change in the relative or total
intensities of the heterodyning longitudinal modes corresponds to a
change in the beat frequency. Thus the uncertainty in the beat frequency
is found by estimating the precision to which we can achieve both the
desired mode relative intensity and desired gain curve
amplitude. Using the Fabry-Perot interferometer, we find that we can steadily hold
the two longitudinal modes at equal relative intensities, resulting in
a negligible uncertainty of $\approx \pm 2$\,kHz due to frequency pulling.
Most of the uncertainty comes from frequency
pushing, for it is not as simple to hold the total amplitude of the
gain curve at a fixed value. To estimate this uncertainty, the
precision to which the amplitudes of the modes can be held constant is
converted into an uncertainty in frequency from the spread of beat
frequencies observed simultaneously on the spectrum analyzer. We
observe that by adjusting the position and aperture size of the iris
in the resonator, we can manipulate the output to have two
longitudinal modes with equal intensity and an overall gain amplitude
that is constant to within $\pm 10\%$. Figure~\ref{fig:pushing}
shows the beat frequency as a function of the total mode amplitude
for our system. A $\pm 10\%$ variation in the total mode amplitude
corresponds to an uncertainty in a single measurement of the beat
frequency of $\pm9$\,kHz.

The uncertainty in the frequency measurement, $\sigma_{\Delta f}$, and the uncertainty in the length measurement, $\sigma_{\Delta L}$, are fixed for each data point, but the uncertainty in the reciprocal beat frequency, $\sigma_{{1/\Delta f}}$, is a function of $\Delta f$ (which varies for each data point). Hence the uncertainty in $1/{\Delta f}$ is not fixed for each data point: $\sigma_{{1/\Delta f}}=\sigma_{\Delta f}/(\Delta f)^2$. Additionally the equivalent uncertainty in $\Delta L$ due to
the uncertainty in $\Delta f$ is of the same order of magnitude as
$\sigma_{\Delta L}$. That is,
\begin{equation}
{{d (\Delta L)} \over{d ({1 \over{\Delta f}})}} \sigma_{1/\Delta f} \approx \sigma_{\Delta L}.
\end{equation}
For this reason, a weighted least squares regression incorporating uncertainty in both variables is performed for the $\Delta L$ versus $1/\Delta f$ data.\cite{Bevington}

The final result for the speed of light in air based on the data plotted in Fig.~6 is
\begin{equation}
c = (2.9972 \pm 0.0002)\times10^8\,\mbox{m/s}.
\end{equation}
The uncertainty of $\pm 0.0002$ is small enough to discriminate
between the speed of light in air ($2.9971 \times 10^8$\,m/s
for $n_{\rm air} = 1.00027$) and the speed of light in a vacuum
($2.9979 \times 10^8$\,m/s).

The measured speed of light yields an index of refraction for air in
our lab of $n_{\rm air} = 1.00024 \pm 0.00006$. We compare this value
to the index of refraction of air as a function of temperature,
wavelength, pressure, and humidity. At conditions of $20^{\circ}$C,
632.8\,nm, 1 atm, and 40\% relative humidity, the
accepted index of refraction of air is 1.00027.\cite{NIST} No
realistic changes in relative humidity, room temperature, or
atmospheric pressure significantly affect the result. Therefore,
the method described here does not have the necessary precision to
demonstrate the effects of atmospheric fluctuations on the index of
refraction.

\section{Conclusion}
This experiment exposes students to a variety of experimental and mathematical techniques, demonstrates the importance of uncertainty in measurement, provides a meaningful context for using weighted regression, and familiarizes the student with three ubiquitous instruments:  the laser, the Fabry-Perot interferometer, and the RF spectrum analyzer.  In addition the experiment yields satisfying results, allowing measurement of the speed of light to a precision which differentiates between the speed of light in air and the speed of light in a vacuum.  The precision to which the measurement is taken is limited by both the precision of our length measurement and our ability to minimize uncertainties due to the frequency pushing and pulling.  One could improve length measurements with a precision linear stage and one could lock the HeNe laser so that the longitudinal modes are held to the same amplitude, but both of these improvements would be beyond the necessary scope of an intermediate physics laboratory course.  

\begin{acknowledgments}
The authors would like to thank all of the Juniata College students who have
performed this measurement in the Advanced Physics Lab over the past
five years. The students that have been particularly instrumental in
improving the experimental technique or data analysis have been
included as authors. We also thank the reviewers for their insightful comments. This work has been supported by the von Liebig
Foundation and NSF PHY-0653518.
\end{acknowledgments}



\begin{thebibliography}{5}
\bibitem[18]{Melles}Kits with HeNe plasma tubes, with one or no internal mirror and windows (either straight or at the Brewster angle) as well as an output coupler and mounts, can be purchesed from Melles Griot.
\end{thebibliography}
\end{document}